\documentclass{article}

\usepackage{PRIMEarxiv}

\usepackage[utf8]{inputenc} 
\usepackage[T1]{fontenc}    
\usepackage{hyperref}       
\usepackage{url}            
\usepackage{booktabs}       
\usepackage{amsfonts}       
\usepackage{nicefrac}       
\usepackage{microtype}      
\usepackage{lipsum}
\usepackage{fancyhdr}       
\usepackage{graphicx}       
\graphicspath{{media/}}     
\newtheorem{theorem}{Theorem}

\newtheorem{definition}{Definition}
\title{zkFaith: Soonami's Zero-Knowledge Identity Protocol
}

\author{
  Mina Namazi \\
  Gmh Soonami.io \\
  Berlin, Germany \\
  \texttt{Mina@soonami.io} \\
   \And
  Duncan Ross  \\
  Gmh Soonami.io \\
  Berlin, Germany\\
  \texttt{Duncan@soonami.io} \\
  \And
  Xiaojie Zhu \\
  Abu Dhabi University \\
  Abu Dhabi, UAE \\ 
  \textit{xiaojie.zhu@adu.ac.ae}\\
  \And
  Erman Ayday \\
  Case Western University\\
  Cleveland, Ohio \\
  \textit{exa208@case.edu}\\
}  
   
\begin{document}
\maketitle

\begin{abstract}
Individuals are encouraged to prove their eligibility to access specific services regularly. However, providing various organizations with personal data spreads sensitive information and endangers people's privacy. Hence, privacy-preserving identification systems that enable individuals to prove they are permitted to use specific services are required to fill the gap. Cryptographic techniques are deployed to construct identity proofs across the internet; nonetheless, they do not offer complete control over personal data or prevent users from forging and submitting fake data. 

In this paper, we design a privacy-preserving identity protocol called "zkFaith." A new approach to obtain a verified zero-knowledge identity unique to each individual. The protocol verifies the integrity of the documents provided by the individuals and issues a zero-knowledge-based id without revealing any information to the authenticator or verifier. The zkFaith leverages an aggregated version of the Camenisch-Lysyanskaya (CL) signature scheme to sign the user's commitment to the verified personal data. Then the users with a zero-knowledge proof system can prove that they own the required attributes of the access criterion of the requested service providers. Vector commitment and their position binding property enables us to, later on, update the commitments based on the modification of the personal data; hence update the issued zkFaith id with no requirement of initiating the protocol from scratch. 
We show that the design and implementation of the zkFaith with the generated proofs in real-world scenarios are scalable and comparable with the state-of-the-art schemes. 
\end{abstract}


\section{Introduction}
Privacy-preserving identification refers to techniques such as pseudonyms, anonymity networks, and secure authentication protocols that allows individuals to identify themselves while protecting their personal information from unauthorized access or disclosure. There are numerous examples in the network where individuals are required to provide access to their documents, photos, and id data to use web services. Dating apps ask you to upload a face photo to verify the real profiles. Local media channels demand location data to allow the viewers to access their content. Age-restricted platforms require to prove that the user is above a certain age. In an elementary example, users should prove that they are human to access some services.

An anonymous credential \cite{chaum1985security,camenisch2002signature,camenisch2004signature,belenkiy2008p, baldimtsi2013anonymous,garman2013decentralized,camenisch2015composable,sonnino2018coconut} is a digital identity that allows users to prove their identity to a relying party without revealing it. It is a privacy-enhancing technology that enables users to authenticate themselves and access services or resources without revealing their personal information. Anonymous credentials are typically based on cryptographic protocols that enable users to prove that they possess specific attributes, such as age or citizenship, without revealing their identity. Anonymous credentials found a vast interest in decentralized platforms where a distributed computation record between many devices known as nodes with no trusted setup is performed. Because the identity of the parties' identity and transmitted data should be hidden, solutions such as Zcash \cite{sasson2014zerocash} are introduced.

Individuals are required to prove their eligibility to access various organizations' services. For example, a streaming platform requires a user's age to be above $18$ and their location in Canada. If users desire to register on this platform and access the content, they must show their ID card information to prove their age and location. A problem appears when the id card contains other information such as nationality, height, eye color, and religion that the streaming platform does not need to see or that the user is unwilling to reveal to the streaming platform. Therefore, a universal identification protocol is required to allow people to prove they are legitimate users without revealing extra information. 

In decentralized platforms, stakeholders deploy cryptographic methods to provide privacy-preserving identification and authentication mechanism. Practical computational integrity proof systems, like SNARKs \cite{groth2016size}, STARKs \cite{ben2018scalable}, and Bulletproofs \cite{bunz2018bulletproofs}, found numerous potential use cases in the recent investigation. These proofs are required not to reveal any additional information beyond the statement of the prover; hence they are called zero-knowledge (zk). In the zk scheme, a prover holds a statement willing to convince the other party, the verifier, of its correctness without revealing any further information. Often, these proof systems require expensive computations. When zk proofs are deployed as a verification system, the proof size must remain small regardless of the computation complexity.

Additional to the scalability challenges of ZK proofs in real-world scenarios, they have no built-in mechanism to prevent individuals from submitting fake data. Current techniques cannot prevent a malicious user who is not eligible to access a streaming platform from using the id card information of another eligible user to gain access.

\noindent \textbf{Our contribution.} This work focuses on developing a zero-knowledge id protocol with data integrity in decentralized platforms called "zkFaith" with the following properties.
\begin{itemize}

\item No trusted setup is required. The issued id perfectly hides the identity of the users and the transmitted data.

\item It provides data integrity since The information inside the credential gets authentication from government-approved authorities. Obtaining a credential for fake data is not possible anymore.

\item It provides privacy to the users. Showing the zkFaith id leaks no data beyond the statement the user is willing to prove. Multiple shows of the same credential to various (or the same) service providers are unlinkable to each other. They reveal no information about the identity or sensitive information of the users.

\item The zkFaith id is updatable. If any document information is modified, the zkFaith id can be updated accordingly upon the user's request with no requirement of initiating the protocol from scratch.  

\item The proposed protocol provides a scalable solution to protect the privacy of individuals and provide data integrity simultaneously. 
\end{itemize}

The rest of this paper is structured as follows. In Section  \ref{related}, we comprehensively review recent developments in identity protocols in which zero-knowledge proofs are used and implemented. In section \ref{prelem}, we represent the primitives and required building blocks to construct our solution. Our proposal is represented in \ref{zkfaith}. We discuss our protocol's security in Section \ref{sec} and the main features and considerations in Section \ref{discuss}. Finally, Section \ref{conclusion} concludes the discussion of the proposed solution and represents open lines of future work.

\section{Related Work}
\label{related}

This section describes the previous work to achieve an efficient zero-knowledge identity. We discuss the advantages and disadvantages of their work and illustrate the differences between our solutions.

\subsection{Coconut}

Sonnino \emph{et al.} \cite{sonnino2018coconut} proposed a credential-based mechanism called Coconut. It is a cryptographic protocol that allows users to share selective information from their credentials (such as a driver's license or passport) with others while keeping the rest of the information private, especially when interacting with distributed ledger systems.
For example, users might want to share proof of their identity with a distributed ledger to access certain services. However, they might want to keep their identity and other sensitive information private. In this case, the user could use Coconut to share a threshold issuance of their credentials, allowing the ledger to verify the authenticity of the credentials without having access to the entire content.
Other contexts, such as online marketplaces or identity verification purposes, are also supported. It provides a way for users to disclose information from their credentials in a privacy-preserving manner selectively. It supports issued credentials by signing under a standard signature (e.g., ECDSA or Schnorr) or by a threshold of parties using a threshold signature scheme. As with other credential systems, data integrity is not supported in Coconut. Moreover, the credentials are issued based on the required criteria of the service providers rather than for general purposes.

\subsection{Deco}

DECO \cite{zhang2020deco} is a decentralized oracle system that allows for secure and transparent data exchange between decentralized applications (DApps) and external data sources. DECO utilizes Transport Layer Security (TLS) to securely exchange data between DApps and external data sources in a decentralized manner, enabling DApps to access and use data from external sources securely and transparently.
One of the main advantages of DECO is that it allows DApps to access external data securely and transparently, enabling them to leverage the data for various purposes, such as making data-driven decisions or providing users with access to external data. 
 Similar to our scenario, a prover commits to personal data and proves to a verifier that this data is obtained from a TLS server. It also generates proof of knowledge of the data. For example, in the proving age scenario, the proof is the predicate $" y/m/d$ is the prover's birth date and the current date - which should be at least 18 years." DECO authenticates the provided information of the prover. The verifier must be convinced that the asserted proof about the data is accurate and that this data is certainly obtained from the website. The protocol is privacy-preserving since the verifier only observes the provided proofs about the data and checks their validity. These proofs leak no information to the verifier about the data or the prover. 

The protocol comprises a three-party handshake phase to establish session keys. In this query execution phase, the prover fetches the server for data, and a proof generation phase where the prover proves the query is well-formed, and the response satisfies the desired condition.

Deco claims to solve the problem of authenticity in zero-knowledge credential protocols. However, the three-party handshake and the deployed secret-sharing schemes seem expensive operations. We achieve authenticity in our proposed solution with no TLS and handshake requirements. In our scenario, an authenticator is a government-approved entity with no access to the parameters to execute the protocol and solely provides authenticity. 

\subsection{zk-cred}

A similar approach to our proposed solution is the zk-cred system introduced in \cite{rosenberg2022texttt}. Usually, the digital platform's users must prove they are legitimate to access the platform. They prove they are human or located in a specific place to access the local services. The zk-cred avoids making unrealistic assumptions where multiple trusted sources exist to issue credentials. It removes the burden of the credential issuers to store various signing keys and proposes a solution using the general zero-knowledge proof system.

A credential in zk-creds is a commitment to arbitrary attributes, such as the fields name or birth date from a passport, placed on a list. The users must convince the credential issuer about possessing it. They must keep a witness to the credential's membership in the issued list and ask the issuer for an updated witness. They deployed a Merkle tree approach to save the list. Therefore, just updating one witness element in the tree is straightforward. The issuer maintains a list of publically verifiable credentials. Anyone can access the list and verify the issue process.   Later, users desire to show their credentials and prove some criteria, such as age. They produce zero-knowledge proof that their credential is listed in a Merkle tree of all issued credentials and meets the access criteria, which is publically verifiable. The revocation mechanism is simply removing the credential from the list. 

A credential is a commitment to some attributes matching the digital identity documents with some random information. They use the Pederson commitment scheme. There is an issuer who keeps the list of credentials in a list as a Merkle tree. The user sends cred to the issuer with supporting documents such as a zero-knowledge proof or a digital signature to convince the issuer of specific criteria. If the issuer is satisfied, it adds the cred to its list and returns the Merkle authentication path. 

A primary difference of zk-cred with our proposed solution is that we assume the authenticator cannot access the zero-knowledge-proof system's public parameters and cannot issue or verify any signature/proof. Moreover, the issuer is not a trusted entity and blindly issues the signature to the users without saving the issued credential information. Also, our scheme provides data integrity before issuing any identity to prevent obtaining an id on submitting fake information.

\subsection{Town Crier}

The town crier \cite{zhang2016town} is similar to Deco's approach to providing a privacy-preserving credential system with a hardware infrastructure. It provides an authenticated data feed (ADF) for smart contracts. TC functions as a high-trust bridge between existing HTTPS-enabled data websites and the Ethereum blockchain. It retrieves website data and serves it to relying contracts on the blockchain as concise pieces of data called datagrams. TC uses a novel combination of Software Guard Extensions (SGX), Intel's recently released trusted hardware capability, and a smart-contract front end. It executes its core functionality as a trusted piece of code in an SGX enclave, which protects against malicious processes and the OS and can attest (prove) to a remote client that the client is interacting with a legitimate, SGX-backed instance of the TC code. This protocol is efficient. However, it depends on the hardware implementation back end and is hard to adopt in real-life scenarios.

\section{Preleminiries and Building Blocks}
\label{prelem}

This section provides the core cryptographic primitives required to develop our zkFaith proposal.

\subsection{Notation.}
\label{notation}
We denote sampling uniformly from a set $S$ by $y \leftarrow S$. Proof of knowledge of a relation $R =
\{(x; w) : P(x, w)\}$ for an instance $x$ is a proof of knowledge of the witness $w$ such that $P(x, w)$ is satisfied, for a predicate $P$. A commitment to a value $x$ with randomness $r$ is $com(x)$. The calligraphic letter $\mathcal{X}$ denotes the parties playing in the protocol. Value $x$ inside brace $[x]$ shows the encrypted version of $x$. Bold letters $\textbf{X}$ denote a vector.

\subsection{Bilinear Maps}
\label{bilinear}

Let $\mathcal{G}$ be a bilinear group generator that on security parameter $k$ returns $(p, \mathbb{G}_1, \mathbb{G}_2, \mathbb{G}_T, e, G, H) \leftarrow \mathcal{G}(1^k)$,  with the following properties: 
\begin{itemize}

\item $\mathbb{G}_1, \mathbb{G}_2, \mathbb{G}_T$ are groups of order $p$;

\item $e = \mathbb{G}_1 \times \mathbb{G}_2 \rightarrow \mathbb{G}_T$ is a non-degenerate bilinear map such that:

\item $\psi = \mathbb{G}_2 \rightarrow \mathbb{G}_1$ is a homomorphism of the form $\psi(H)=G$, hence $\psi(H^a)=G^a,~ \forall a \in \mathbb{Z}$

\item $ \forall, V, U \in G, ~\forall a, b \in Z : e(aU, bV ) = e(U, V )^{ab}$.

\item $G$ generates $\mathbb{G}_1$, $H$ generates $\mathbb{G}_2$ and $e(G, H)$ generates $\mathbb{G}_T$.
\end{itemize}

\subsection{Non-Interactive Zero-Knowledge Proof of Knowledge (NIZK)}
\label{nizk}

Groth \cite{groth2016size} introduced a proof system with the following functionalities.

\begin{itemize}
    \item $\mathrm{NIZK.Setup}(1^\lambda) \rightarrow pp$: generates public parameters for the bilinear group. It is the input to all other algorithms.
    
\item $\mathrm{NIZK.Prove}( x, w) \rightarrow \pi$:  generates a proof, where $x$ is the statement, and $w$ is the witness.

\item $\mathrm{NIZK.Verify}( \pi, x) \rightarrow \{0, 1\}$: verifies the proof $\pi$ with respect to $x$ and returns $1$ as a truthy value.
\end{itemize}



\subsection{Vector Commitments}
\label{vectorcommit}

A VC is a position-binding commitment and can be opened at any position to a unique value with short proof. Boneh \emph{et al.} introduced vector commitment (VC) used as a communication-efficient authenticated data structure (ADS) for a remotely stored database where users can retrieve individual items along with their membership proofs in the data structure. They have constant size openings and public parameters, therefore suitable to our protocol. This VC can be hiding as well and comprised of the four algorithms: $\mathrm{VC.Setup}$,
$\mathrm{VC.Com}$, $\mathrm{VC.Open}$, $\mathrm{VC.Verify}$ described as follows.
\begin{itemize}

\item $\mathrm{VC.Setup}(\lambda, n,M) \rightarrow pp$. It inputs the security parameter $\lambda$, length $n$ of the vector, and message space of vector components $M$, outputs public parameters $pp_{VC}$. The vector commitments are position-binding as their main security property.

\item $\mathrm{VC.Com}(M) \rightarrow (\tau, com)$. It inputs $M = (m^{(1)}, \ldots, m^{(l)})$ output a commitment $com$ and advice $\tau$ .

\item $\mathrm{VC.Update}(com, M, i, \tau ) \rightarrow (\tau, com)$. It inputs message $m$ and position $i$. It outputs a commitment $com$ and advice $\tau$.

\item $\mathrm{VC.Open}(com, M, i, \tau ) \rightarrow \pi$. It inputs $m \in M$ and $i \in [1, n]$, the commitment
$com$, and advice $\tau$ output an opening $\pi$ that proves $m$ is the $i$'th committed
element of $com$.
\item $\mathrm{VC.Verify}(com, m, i, \pi) \rightarrow 0/1$. It inputs commitment $com$, an index $i \in [n]$,
and an opening proof $\pi$ output $1$ (accept) or $0$ (reject).
\end{itemize}

\subsection{CL Signatures with Aggregation Features}
\label{clsig}

A digital signature scheme is a way of signing documents and a functional building block to construct an anonymous identification protocol. In this paper, individuals must obtain a signature on their confidential information to build an efficient identity-based solution. Lee \emph{et al.} \cite{lee2013aggregating} proved that the Camenisch-Lysyanskaya (CL) signature scheme \cite{camenisch2004signature} can be an aggregated signature scheme. Hence, different signatures from various signers (or the same signer on various messages) can be compacted into one signature scheme with a succinct verification process. We embed vector commitment explained in section \ref{vectorcommit} into the aggregated CL signature to obtain a signature on a vector of commitments and deploy it to construct the proposed zero-knowledge identity protocol. Later, we explain how to use this scheme to prove the knowledge of this signature on their information in a zero-knowledge way.

$\mathrm{CL.Setup}$: inputs the security parameter $1^\lambda$ and outputs $pp_{CL} = (\mathbb{G},G, \textbf{g}, g, e)$. These parameters are inputs of all the other algorithms.

 $\mathrm{CL.KeyGen}$: each user does the following 

\begin{itemize}

\item Choose $x \leftarrow Z_q, y \leftarrow Z_q$, and for $1 \leq i \leq l, z_i \leftarrow Z_q$. 
\item Let $X = g^x, Y = g^y$ and, for $1 \leq i \leq l, Z_i = g^{z_i}$, $W_i = Y^{z_i}$. 

\item Return $sk = (x, y, z_1, \ldots, z_l)$, and $pk = (q, \mathbb{G},G, \textbf{g}, g, e, X, Y, \{Z_i\}, \{W_i\})$.

\end{itemize}

 $\mathrm{CL.AskSig}$: the algorithm inputs the committed message 
 $M =(m^{(0)}, m^{(1)}, \ldots , m^{(l)})$, signer's secret key $sk = (x, y, z_1, \ldots, z_l)$, and public key $pk = (q, \mathbb{G},G, g, \textbf{g}, e, X, Y, \{ Z_i \}, \{ W_i \})$. Then it continues as follows.

\begin{itemize}

    \item  The user calls $\mathrm{VC.Com}(M)$ to generate $com$ and sends proof of knowledge of the commitment's opening to the signature issuer $\pi_i$. 
  
 \end{itemize}
 
 $\mathrm{CL.IssueSig}$: The signature issuer runs the algorithm and calls $\mathrm{VC.Verify}(com, m, i, \pi)$. If it is satisfied with the proof of knowledge of the commitment opening acts as follows.
 
 \begin{itemize}

  \item  The issuer chooses a random $\alpha \rightarrow \mathbb{Z}_q$, calculates $a = g^{\alpha}$. Let $A_i = a^{z_{i}}$, for $1 \leq i \leq l$, let $b = a^y, B_i = (A_i)^y$. Let $c = a^x com(M)^{\alpha x y}$. 
  \item The user outputs the signature as $\sigma = (a, \{ A_i \}, b, \{ B_i \}, c)$.
\end{itemize}

\begin{sloppypar}
$\mathrm{CL.Verify}$: the algorithm is run by the user to verify the correctness of the signature. It inputs $pk = (q, \mathbb{G},G, g, \textbf{g}, e, X, Y, {Z_i})$, message $M$, and signature $\sigma = (a, {A_i}, b, {B_i}, c)$, and checks as it follows.
\end{sloppypar}

\begin{itemize}
    \item  $\{A_i\}$ were formed correctly: $e(a, Z_i) = e(g, A_i)$.

\item $b$ and $\{B_i\}$ were formed correctly: $e(a, Y ) = e(g, b)$ and $e(A_i, Y ) = e(g, B_i)$.

\item $c$ was formed correctly: $e(X, a) \cdot e(X, b)^{m^{(0)}} \Pi_{i=1}^l e(X, B_i)^{m^{(i)}} = e(g, c)$.

\end{itemize}

This signature scheme is a secure two-party computation of a signature on a discrete logarithm representation of the message $com(M)$ under the signer's public key. The signer can only see a commitment to an array of messages where the commitment scheme is hiding and does not reveal any information about the user's raw data to the signer. 

A signature scheme is \textbf{malleable} if, on the input of a message $m$ and a signature $\sigma$, efficiently computing a signature $\sigma^\prime$ on a related message $m^\prime$ is possible, $m^\prime = T (m)$, for a permitted transformation $T$. Chase \emph{et al.} \cite{chase2014malleable} introduced a malleable signature that benefits our protocol. A user can obtain different unlinkable zk identities for their various contexts. Then they can prove possession of an issued id to one document to a verifier by showing another record with updated information. Let $T$ be the set of transformations that, on the input of some legal data, output another data of the same user, i.e., for every such pair $doc, doc^\prime$, there exists some $T$ such that $doc^\prime = T(doc)$. Then a malleable signature scheme concerning $T$ achieves a zero-knowledge identity system. A zk identity is an authority's signature $\sigma$ on $doc$. Malleability empowers users to transform the currently issued signature into another signature $\sigma^\prime$ on $doc^\prime$. Context hiding ensures that $\sigma^\prime$ cannot be linked to the original document, and unforgeability ensures that users cannot compute $\sigma^\prime$ unless they receive a signature from the issuer on one of their government-issued documents. The following function transfers the signature $\sigma$ to $\sigma^\prime$ with respect to tranformation $T$:

$\mathrm{CL.SigEval}(crs, vk, T, m, \sigma)$: Set $T_{inst}$ to be such
that $\forall m,vk, T_{inst}(vk,m) = (vk,T(m))$, and $T_{wit} = id$; i.e., such that $\forall sk , T_{wit}(sk ) = sk$. Return $\sigma \leftarrow \mathrm{mZKEval}(crs, (T_{inst}, T_{wit}), (vk, m), \sigma)$.

\section{The Proposed Solution: zkFaith }
\label{zkfaith}

This section describes our proposal for a zero-knowledge identification and authentication system of "zkFaith." First, we describe the core cryptographic primitives and building blocks to construct our proposed solution. Then, we explain the zkFaith protocol and its deployment in real-life scenarios. Finally, we discuss the protocol's security goals and design challenges in addition to its complexity. 

Informally, the users obtain a zero-knowledge identity after a third-party government-approved authority confirms their provided information. Each party can obtain various identities on their different legal documents. However, a unique identification number, such as each party's passport or wallet number, is a shared value in all identities. The parties can prove that they own a legitimate identity based on their information with zero knowledge. The issued id is updatable. Therefore, if a passport is expired, the user can ask to update their zero-knowledge identity. The zkFaith is unique, and there is no way that a user is issued two different identities for the same information. In case of malicious activity, such as sharing the credential, the deployed revocation mechanism removes the issued credential and is no longer functional. We explain the details of the protocol in the following sections.

 \subsection{System Model}
\label{model}

The protocol comprises a claimant $\mathcal{C}$ who asks for a zkFaith identity by submitting an official document. The government previously issued these documents, such as a passport, driving license, or medical certificate. We denote the document by $doc$, and $info_i$ refers to the information inside each document, such as name, sure-name, birth date, expiry, or nationality.   

An authority $\mathcal{A}$ is a government-approved organization that can verify the authenticity and integrity of the data submitted by $\mathcal{C}$. If $\mathcal{A}$ is convinced that $\mathcal{C}$  submitted a valid document of its own, it sends an authentication tag to the zkFaith issuer entity.

There is an issuer $\mathcal{I}$, in our scenario Soonami (or unbounded DAO), who receives the authentication tag from the $\mathcal{A}$ and issues the zero-knowledge identity of zkFaith to $\mathcal{C}$.

Some verifiers $\mathcal{V}$ such as websites, shops, and organizations within some defined criteria $\phi$ receive the zkFaith id of the users and decide to grant access to the users.

The zkFaith is a protocol that aims to create compliant data proof for identity, medical data, rights, membership, and other forms of personal data stored in a secure wallet with on-chain verification and data integrity check by a recognized authority, therefore providing real-world compliance, a form of KYC and usability without compromising or revealing user data.

Each identity is not self-sovereign and can not issue a claim on another identity (person, organization, or system/machine), preserving compliance and verifiability of the claimed data. The protocol is zero-knowledge; therefore, the identity of the parties and transmitted data reveal no information about the parties or the transactions.

The $\mathcal{C}$ starts the protocol by sending their documents to the authority $\mathcal{A}$, who checks their validity. If $\mathcal{A}$ is convinced that the submitted documents are valid, it generates a tag and sends it to the issuer $\mathcal{I}$. Then, $\mathcal{C}$ commits to the verified documents and sends the commitment and proof of an opening to the $\mathcal{I}$. The issuer checks the proofs and signs the claimant's commitment with their private key. Then they generate proof of the correctness of the signature and send them to the claimant, a unique zero-knowledge identity for each user. The issuer also creates a list of revoked zkFaith ids. Each user is required to generate a non-membership proof of this list.

This procedure is a one-time operation. Later, suppose the users desire to prove their eligibility for accessing a specific web service in which the only requirement is proving the user is above $18$ years old. In that case, the verifier must check a subproof of knowledge of the user's birth year and the signature's validity and attach it to the proof of valid zkFaith identity. If the website can verify the proofs, it grants access to the user and allows them to benefit from their services. 

\begin{figure}
  \centering
  \includegraphics [scale=0.4]{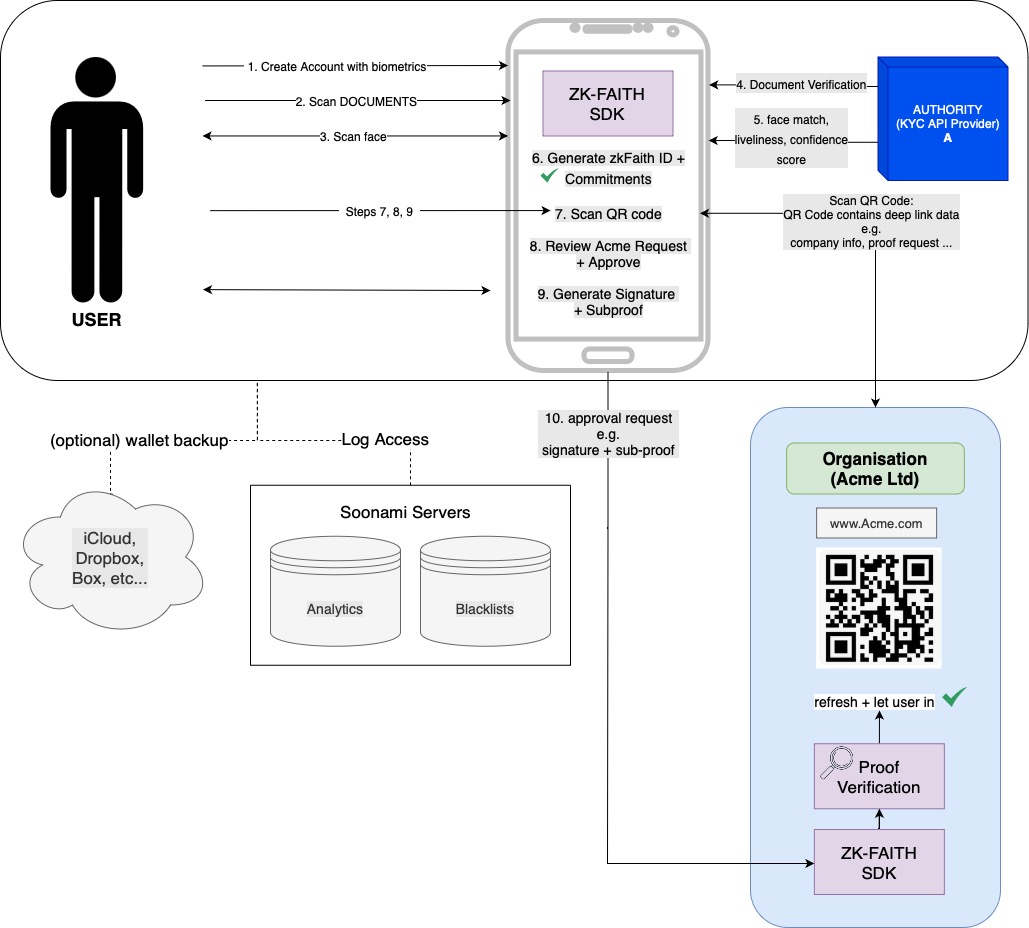}
  \caption{The Interactions of the Claimant with KYC provider, Issuer, and Verifier in zkFaith Protocol.}
  \label{fig:fig1}
\end{figure}

If information inside the zkFaith id is changed, such as passport expiry date, the claimant sends the commitment to the new information and asks the issuer to update their zkFaith id. We explain the interactions between parties in Figure \ref{fig:fig1}.

\subsection{Security Goal and Threat Model}
\label{secgoal}

The proposed solution's security goal is to protect the privacy of individuals and provide data integrity. Privacy means that in any step of the protocol, any adversarial behavior cannot bind the result of the protocol interactions to any specific claimant. Various deployments of the zkFaith Id are unlinkable to each other and reveal nobody's identity. We assume that the authenticator, $\mathcal{A}$,  is a trusted party with no access to the system parameters, and no information storage related to the individuals after authenticating them is permitted.

The authority $\mathcal{A}$ is a trusted party. However, we assume that it only controls the validity of the provided document and does not store them. In our scenario, the claimant $\mathcal{C}$ can be controlled by bounded malicious adversaries who are allowed to alter the protocol instructions and seek to modify the inputs to statically corrupt the system among the instances that our security goals address. The issuer is semi-honest, follows the protocols' instructions, and cannot modify the data. However, it can be curious and get unauthorized information from the interactions and outputs of the protocol. The verifiers can be malicious and interact with each user to steal their zkFaith id or identify the users.

The authenticator and issuer in real-life scenarios are not colluding because such a colliding will endanger the reputation of both organizations. 


\subsection{Design Challenges}
\label{challenge}
This section lists potential attack scenarios a malicious adversary may deem to attempt in our system model. We also outline our proposed solution for each scenario. While the attacks are not confined to the listed, we emphasize the design thinking that would overcome the security challenges by determining an intuition of the cryptographic primitives.

One possible attack scenario is when a malicious user (hereafter known as "attacker") generates fake profiles using the identification information of other honest users (hereafter known as "claimant"). Let us assume that a claimant obtains a zkFaith identity on their passport. The attacker, who owns no driving license, takes control of the claimant's document and submits it to ask the issuer for a zkFaith on the driving license by committing to their driving license information and the malicious user's public key. If the issuer is convinced with the claimant's proof, it issues the zkFaith for the malicious user. We defeat this attack by requiring an authentication step before issuing the zkFaith. The claimants and the validity of the represented document should be authenticated. After a tag is generated and sent to the issuer, it continues the protocol. Otherwise, no new zkFaith for the provided driving license is generated, and the protocol halts.
 
Another possible scenario is that the issuer issues an identity on arbitrary messages not provided by the claimant and uses the claimant to access some services through them. The protocol prevents the attack by requiring the claimant to check the issued signature by the issuer for the "correct computation" on the provided data. The protocol aborts if the generated CL signature by the issuer does not satisfy the verification step of the claimant.

Moreover, when an organization with some criteria to allow the users to access their services requires the claimants to prove that they meet their required criteria. They can relate two shown credentials to each other and trace them back to identify the individuals. We defeat this attack by deploying a zero-knowledge-proof system that is randomizable. Various shows of the zkFaith on the same document or two zkFaith issued for different documents of the same claimant are not linkable. 

Last but not least, an attacker can ask for various subproofs and combines them to re-create the zkFaith Id of the claimant and abuse it for its own malicious purposes. The shown proofs are random points in the field, and combining leads to no meaningful proof for the same statement.

\subsection{zkFaith Definition}
\label{def}
This section combines all the preliminaries and constructs a zkFaith protocol. The zkFaith protocol comprises of $Faith =  \{ \mathrm{Faith.Setup}, \mathrm{Faith.KeyGen}, \mathrm{Faith.Auth}, \mathrm{Faith.Ask}, \mathrm{Faith. Issue}, \mathrm{Faith. Show}, \mathrm{Faith. Update}, \mathrm{Faith. Revoke} \}$.  The definition of each step is explained as follows. 

We assume that each user is pre-issued an identity document (without loss of generality, hereafter, we denote it by passport.) by the government. They receive a document $doc$ with \textit{inf}$_i$ to be each section of the $doc$ kept as a secret. Moreover, each user possesses a unique wallet id, $wid$.

\noindent \textbf{Set up and Key Generation:}

\begin{itemize}

\item $pp_F \leftarrow \mathrm{Faith.Setup}(1^\lambda)$: The algorithm takes a security parameter $\lambda$ and outputs the system parameters $pp_{F}$ that are inputs to all of the below algorithms.

\item  $(sk_u, pk_u) \leftarrow \mathrm{Faith.KeyGen}(pp_F)$: Each party $\mathcal{U}$ on the input $pp_{F}$ generates a private and public key pairs. 

\end{itemize}

\noindent \textbf{Request:}
\begin{itemize}

\item $R \leftarrow \mathrm{Faith.Auth}(wid, doc)$: $\mathcal{C}$ sends previously government-issued ${doc}$ to the authority. Then, $\mathcal{A}$ outputs a response $R = (wid||1)$ to $\mathcal{I}$. Otherwise, it aborts the protocol.

\end{itemize}

\noindent \textbf{Issue:}

\begin{itemize}

\item $Q \leftarrow \mathrm{Faith.Ask}(com(doc)$: $\mathcal{C}$ sends the commitments to the information of the claimant's $com(doc)$ and outputs $\pi_i$s on the given identity information.

\item $(\sigma, \pi_{\sigma}, L) \leftarrow \mathrm{Issue}(R, Q)$: The issuer inputs the authentication response $R$ from $\mathcal{A}$ for the corresponding wallet id of the claimant and the request $Q$ from the $\mathcal{C}$ and outputs $\sigma$ with its correctness proof $\pi_{\sigma}$. 

\item The issuer generates an empty list $L$ of revoked zkFaith ids. The claimant is required to prove that their id does not belong to this list using a non-membership proof $\pi_{no-mem}$.

\item If $\mathcal{C}$ is convinced that the signature is correctly generated on the previously provided data, it outputs zkFaith id, and aborts otherwise.

\end{itemize}

\noindent \textbf{Show:}

\begin{itemize}

\item $(0,1) \leftarrow \mathrm{Faith.Show}(\pi_{no-mem}, \pi_\sigma, com(M))$:  The verifier $\mathcal{V}$ with respect to the defined criteria $\phi$ receives  a commitment to the related message, proof of a knowledge of a signature and proof of valid zkFaith id. It outputs $1$ and grants access to the claimant. Otherwise, aborts the protocol.
\end{itemize}
\noindent \textbf{Update:}

\begin{itemize}

\item  $(\sigma^\prime, \pi_{\sigma^\prime}) \leftarrow \mathrm{Faith.Update}(Q, Q^\prime)$: It inputs the user's current query $Q$ and updated query $Q^\prime$ and transforms the signature $\sigma$ to $\sigma^\prime$ with its proof of correctness $\pi_{\sigma^\prime}$.
\end{itemize}
\noindent \textbf{Revoke:}

 In case of any malicious activity, the issuer adds the zkFaith id into the revoked id list $L$.


\subsection{zkFaith Instantiation}
\label{instant}
 
This section breaks down the proposal introduced section~\ref{def} and explains interactions of separate protocols in detail. Before initializing the protocol, we assume that individuals receive an id from the government, such as a passport number. This id number is unique and belongs to each user. Moreover, this id number is the common value that each user is required to include while asking for data authentication from the $\mathcal{A}$.

\noindent \textbf{SetUp and Key Generation:}
\begin{itemize}

\item $pp_F \leftarrow \mathrm{Faith.Setup}(1^\lambda)$: The algorithm takes a security parameter $\lambda$, calls the $\mathrm{CL.Setup}$ and outputs the system parameters $pp_{CL}$. It assigns $pp_F = pp_{CL}$ and inputs them to all of the below algorithms. Let $\mathcal{G}$ be a bilinear group generator that on security parameter $k$ returns $(\mathbb{G},G, \textbf{g}, g, e) \leftarrow \mathcal{G}(1^k)$. These parameters are the input of all other algorithms. The setup process is a pre-computed phase, and the parameters are already distributed between the parties.

\end{itemize}

\noindent \textbf{Request:}

In this phase, the claimant is required to upload a copy of the official document to the KYC provider. The $wid$ is a public wallet id unique for each user.

\begin{itemize}

\item  $(sk_{u}, pk_u) \leftarrow \mathrm{Faith.KeyGen}(pp_F)$: Each party $\mathcal{U}$ on the input $pp_{F}$ calls the key generation algorithm of the CL signature $\mathrm{CL.KeyGen}$ locally to generate a key pairs for each user. Choose $x \leftarrow Z_q, y \leftarrow Z_q$, and for $1 \leq i \leq l, z_i \leftarrow Z_q$. Let $X = g^x, Y = g^y$ and, for $1 \leq i \leq l, Z_i = g^{z_i}$, $W_i = Y^{z_i}$.  Return $sk = (x, y, z_1, \ldots, z_l)$, and $pk = (q, \mathbb{G},G, \textbf{g}, g, e, X, Y, \{Z_i\}, \{W_i\})$.

\item $R \leftarrow \mathrm{Faith.Auth}(wid, doc)$: $\mathcal{C}$ sends previously government-issued ${doc}$ such as passport, driving license which contains information such as name, nationality, expiry day, address represented as $\{inf^{(0)}, inf^{(1)}, inf^{(2)}, \ldots, inf^{(l)} \}$ (coded to binary representation) to the authority. We assume that $\mathcal{A}$ deploys a mechanism such as a face recognition to control that the represented information is correct and belongs to the $\mathcal{C}$. Then, $\mathcal{A}$ outputs a response $R = (wid||1)$ to $\mathcal{I}$. Otherwise, it aborts the protocol.

\item $Q \leftarrow  \mathrm{Faith.Ask}(com(doc))$: $\mathcal{C}$ takes the $inf_{(i)}$ from $doc$ and calles  $\mathrm{CL.AskSig}$. First it builds the vector $M = (m^{(0)}, m^{(1)}, \ldots , m^{(l)}) = (inf^{(0)}, inf^{(1)}, \ldots, \inf^{(l)}) $ constructs the commitment to this vector $M$ such that $com(M) = \Pi_{i=1}^l Z^{m^{(i)}}_i$ by calling $\mathrm{VC.Com}(M) \rightarrow com(M$. It also generates proof of knowledge of $M$ as $\pi_M$ and outputs $Q = (com(M),\pi_M)$


\end{itemize}

\noindent \textbf{Issue:}

\begin{itemize}

\item $(\sigma, \pi_{\sigma}, L) \leftarrow \mathrm{Issue}(R, Q)$: The issuer inputs the authentication response from $\mathcal{A}$ for the corresponding wallet id of the claimant excluding is $wid$ and $1$ parameters. It also receives the request query $Q$ from the $\mathcal{C}$.  Then, $\mathcal{I}$ calls $\mathrm{VC.Verify}(Q)$ and outputs $1$ if satisfied with the proof of knowledge of the commitment. Then the issuer calls the $\mathrm{CL.IssueSig}(sk_{\mathcal{I}}, com(M), wid)$ and outputs $\sigma$ for the claimant. 

\item The issuer is not storing any credentials in their databases. Instead, there is an empty list $L$ that is supposed to be filled with revoked credentials. Each claimant is also asked to send proof of non-membership in the issuer's list $\pi_{no-mem}$.

\item $\mathcal{C}$ calls the signature verification algorithm $ \mathrm{CL.Verify}(pk_{\mathcal{I}}, \sigma, com(M), wid) \rightarrow 0/1$. It takes as inputs the signature $\sigma$, the public key of the issuer, and the $com(M)$ and the claimant's wallet id $wid$. If $\mathcal{C}$ is convinced that the signature is correctly generated on the previously provided data, it outputs $1$, and $0$ otherwise.

The claimant generates a commitment to this signature and a proof of knowledge of the signature. It outputs zkFaith$ = (Com(\sigma), \pi_{\sigma})$ is the unique zero-knowledge identity of the user for the provided documentation of the claimant.

\end{itemize}

\noindent \textbf{Show:}

$(0,1) \leftarrow \mathrm{Faith.Show}(\pi_{no-mem}, \pi_\sigma, com(M))$: The $\mathcal{C}$ is required to prove that the data inside the issued zkFaith id (i.e., $inf_i$'s) satisfies the required condition $\phi$ as it follows.
\begin{itemize}

\item The claimant runs $\mathrm{CL.Prove}({zkFaith})$ to generate the proof of knowledge of the signature $\pi_{\sigma}$ on their committed vector $com(M)$.

\item The claimant also needs to prove possession of a valid zkFaith. i.e., a non-membership proof to show that the zkFaith does not exists in the  list $L$, denoted by $\pi_{no-mem}$.

\item The claimant outputs $(\pi_{no-mem}, \pi_{\sigma}, com(M))$ and shows it to the $\mathcal{V}$ as a proof of knowledge of the correct criteria and proof of knowledge of a valid credential from an issuer.

\end{itemize}
 \textbf{Update:}

If the information inside the document of the claimant is changed from $m_i$ to $m_j$, where $i,j$ are the positions of the information in vector $M$, the claimant performs the following actions requesting to update the previously issued zkFaith id.

\begin{itemize}

\item $(\sigma^\prime, \pi_{\sigma^\prime}) \leftarrow \mathrm{Faith.Update}(Q, Q^\prime)$: It inputs the user's current query $Q$ and updates it to the new query $Q^\prime$. It takes the new message $m_j$ and its position $j$ in vector $M$. It generates the commitment to the new message $com(m_j)$ and calls $\mathrm{VC.Update}(Q, com(m_j), j)$ to update $com(M)$ to $com(M^\prime)$.

\item It also calls $\mathrm{VC.Open}(com (M^\prime), \pi_{M^\prime}, m_j, j ) \rightarrow \psi$ to generate a proof that $m_j$ is in position $j$ in vector $com(M^\prime)$.

\item It returns $Q^\prime = (com (M^\prime), \pi_{M^\prime}, \psi)$.

\item It constructs the allowed transformation $T_{Q \rightarrow Q^\prime}$ and sends it to the issuer.

\item The issuer $\mathcal{I}$ calls $\mathrm{mSigEval}( T, Q, Q^\prime)$ and outputs the zkFaith id as $(\sigma^\prime, \pi_{\sigma^\prime})$.

\end{itemize}

\noindent \textbf{Revoke:}

Revocation of zkFaith is straightforward. In case of malicious activity, the issuer adds the zkFaith id to the revoked id list $L$. The claimant can no longer prove possession of a valid zkFaith id to the verifiers.

The zkFaith protocol we introduced in this section is a privacy-preserving identification protocol that deploys non-interactive, efficient zero-knowledge proofs to prevent data leakage to the issuers or verifiers. The choice of hiding and position-binding vector commitments prevents users from changing their statements and positions after claiming them. The CL signature and its structure-preserving properties with embedded VC enable the protocol to create a zero-knowledge id for the users. It can choose various information from the issued id depending on the criteria of the verifiers. It suffices to install the zkFaith application on the phone, and after that, the protocol automatically generates the zero-knowledge proofs. Hence, the zkFaith is user-friendly, and it is more probable that the users will adopt it in real-life scenarios. 

The protocol is application specific. Depending on the criteria of the verifier, it chooses to prove partial information, such as age, while accessing age-limited content. In scenarios such as registering a company, the protocol chooses to prove the possession of a correct zkFaith id over all the information of the submitted documents. Therefore, it decreases the burden of the users to perform any extra steps while proving their eligibility to various service providers. We discuss the practicality of the zkFaith and its comparison with the state-of-the-art solution in the following section.


\section{Security}
\label{sec}

The protocol introduced in section \ref{def} protects the privacy of individuals when they request a zkFaith id. Moreover, data integrity and authenticity are also provided by the deployment of this solution. We analyze the security of zkFaith against an active static adversary who controls some of the claimants and verifiers. We consider a malicious adversary who neglects the protocols’ instructions and acts arbitrarily and a semi-honest adversary for the issuer. This section presents the security definition of user privacy, and we provide complete security proof in Section \ref{sec}. 

The following game illustrates the formal definition of user privacy. An adversary plays the role of all malicious entities in the system, i.e., some claimants, the issuers, and some verifiers. However, at least one verifier is assumed to be honest. On the other hand, the challenger controls the honest parties, such as KYC providers. If an adversary can convince a verifier with zkFaith id generated on documents that the claimant never submitted, it wins the game. 

Note that this security definition perfectly copes with the security challenges we discussed in Section \ref{challenge}. 

User privacy experiment $UPriB(\lambda)$:
\begin{enumerate}

\item  Adversary and challenger are given the security parameter $\lambda$. They execute "setUp and key generation" protocol, and the adversary receives public parameters and public keys.

\item The adversary plays the role of the claimant. It chooses a $doc^\prime$, submits it to the "request" protocol, and receives the response $R^\prime$ from the challenger, who plays as the KYC.

\item The adversary submits a query $Q^\prime$ and asks the challenger to run the "issue" protocol upon receiving the response, runs the "update" internally, and outputs zkFaith${^\prime}$. These steps can be run polynomially many times.

\item The adversary, who is the claimant now, asks the challenger to run the "show" protocol by submitting zkFaith${^\prime}$. The challenger plays the verifier role and decides on outputting $0$ or $1$.  

\item If the output of the "show" protocol is $1$, it means the adversary can generate a zkFaith id on unauthorized data or can forge a signature and succeed with this game. 


\end{enumerate}

We consider $UPrivB(\lambda)$ a probabilistic experiment defined in terms of a game played between adversary $\mathcal{A}$ and a challenger.

\begin{definition}
\label{def1}
The zkFaith protocol protects user privacy against an active adversary if for every PPT adversary $\mathcal{B}$, there exists a negligible function $negl(\lambda)$, where $\lambda$ is the security parameter, such that: $$Pr[UPrivB(\lambda) = 1] \leq \frac{1}{2} + negl(\lambda)$$
\end{definition}

\textbf{Proof Sketch}: The security proof comprises two parts. Firstly, as illustrated in Theorem \ref{theorem1}, we show that breaking the user privacy of zkFaith compromise the security of the underlying commitment and signature schemes. Through the construction of a simulator $\mathcal{D}$. The second part of the proof is supported in Theorem \ref{theorem2} in which we prove that $\mathcal{D}$ runs indistinguishable from the real challenger due to the randomizeability of the deployed NIZK proofs. 

\begin{theorem}
\label{theorem1}
The zkFaith protocol defined in section \ref{def} is secure and provides user privacy, if the underlying vector commitment \ref{vectorcommit} is hiding and position binding and the CL signature scheme \ref{clsig} is unforgable. 
\end{theorem}

If there exists a PPT adversary $B$ who breaks user privacy with $\epsilon(\lambda)$ advantage, then we can construct a PPT adversary $D$ who breaks the security of the CL signature and hiding properties of VC with the same advantage

For the second part of the proof, we construct a modified simulator $\mathcal{D}^\prime$ which runs identical to $\mathcal{D}$. We prove that $\mathcal{B}$ cannot distinguish interaction with $\mathcal{D}$ and $\mathcal{D}^\prime$ unless the underlying proof scheme is not randomizable. Let $UPriB,D^\prime(\lambda)$ indicate the user privacy experiment run between $\mathcal{B}$ and $\mathcal{D}^\prime$.

\begin{theorem}
\label{theorem2}
 We assume that the NIZK proofs introduced in \ref{nizk} are randomizable, then $$Pr[UPriB,D(\lambda) = 1] - Pr[UPriB,D^\prime(\lambda) = 1]| < negl(\lambda)]$$
\end{theorem}

Then we can construct an adversary who can distinguish between two generated NIZK-proof schemes. The proof is straightforward and directly reduces to indistinguishability properties of the underlying zero-knowledge proof scheme.

\section{Discussion}
\label{discuss}

The protocol introduced in section~\ref{zkfaith} perfectly protects individuals' identities in decentralized platforms. The authority $\mathcal{A}$ is a trusted party. However, it has no access to the common reference string generated during the protocol and parameters of the zero-knowledge protocol. Moreover, it will not store any information related to the parties. The "request" protocol leaks no information about the identity or raw data of the claimant since the commitment scheme is perfectly hiding. In real-world scenarios, the issuer is an automated smart contract. Its procedures are pre-defined. Hence there is no room for collusion between authority and the issuer contract. Later on, when the zkFaith is deployed to access various services during the "show" phase, it leaks no information to the service provider. Since the identity and generated proofs are randomizable, different usage of it is unlikable to each other. The service providers cannot link two shown zkFaiths of the same claimant to each other and reveal the claimant's identity. The relevant authority must have issued the data (i.e., Passport Office, Police, Vehicle Agency, Medical board). Hence, data schema and its format are known. Using CL signatures guarantees data integrity since the data structure is preserved in the signature/proof generation process.


\section{Conclusion and Future Work}
\label{conclusion}

We designed a zero-knowledge identification protocol, zkFaith, and developed its verification procedure. We showed that the devised solution could be used as an anonymous credential to access various organizations' services with specific criteria. We represented that it is a privacy-preserving protocol that simultaneously provides integrity and authenticity of the data. We showed with no trusted setup how our solution could be implemented in decentralized platforms and work with smart contracts with minimum gas fees. 

We defined compliance. Each user can create a transaction deploying the zkFaith protocol and prove its correctness and eligibility for making the transaction with zero knowledge to a potential provider. The transactions and the proofs perfectly hide the origin, destination, and values inside the transactions against a malicious issuer. 

Moreover, an "update" protocol is added to the scheme for scenarios where some parts of the previously issued document are changed, such as expiry dates. The protocol dynamically updates the altered information inside the zkFaith with no requirement of initializing the entire system from scratch by deploying malleable signature schemes. 

Finally, to increase the level of security and privacy in the scheme, an SDK will be embedded in the authentication part to allow the authenticator access to the system parameters and issue/verify commitments or signatures. This feature removed the need for the authenticators to be trusted and access the raw data of the claimants.


\bibliographystyle{unsrt}  
\bibliography{references}

\begin{thebibliography}{10}

\bibitem{chaum1985security}
David Chaum.
\newblock Security without identification: Transaction systems to make big
  brother obsolete.
\newblock {\em Communications of the ACM}, 28(10):1030--1044, 1985.

\bibitem{camenisch2002signature}
Jan Camenisch and Anna Lysyanskaya.
\newblock A signature scheme with efficient protocols.
\newblock In {\em International Conference on Security in Communication
  Networks}, pages 268--289. Springer, 2002.

\bibitem{camenisch2004signature}
Jan Camenisch and Anna Lysyanskaya.
\newblock Signature schemes and anonymous credentials from bilinear maps.
\newblock In {\em Annual international cryptology conference}, pages 56--72.
  Springer, 2004.

\bibitem{belenkiy2008p}
Mira Belenkiy, Melissa Chase, Markulf Kohlweiss, and Anna Lysyanskaya.
\newblock P-signatures and noninteractive anonymous credentials.
\newblock In {\em Theory of Cryptography Conference}, pages 356--374. Springer,
  2008.

\bibitem{baldimtsi2013anonymous}
Foteini Baldimtsi and Anna Lysyanskaya.
\newblock Anonymous credentials light.
\newblock In {\em Proceedings of the 2013 ACM SIGSAC conference on Computer \&
  communications security}, pages 1087--1098, 2013.

\bibitem{garman2013decentralized}
Christina Garman, Matthew Green, and Ian Miers.
\newblock Decentralized anonymous credentials.
\newblock {\em Cryptology ePrint Archive}, 2013.

\bibitem{camenisch2015composable}
Jan Camenisch, Maria Dubovitskaya, Kristiyan Haralambiev, and Markulf
  Kohlweiss.
\newblock Composable and modular anonymous credentials: definitions and
  practical constructions.
\newblock In {\em International Conference on the Theory and Application of
  Cryptology and Information Security}, pages 262--288. Springer, 2015.

\bibitem{sonnino2018coconut}
Alberto Sonnino, Mustafa Al-Bassam, Shehar Bano, Sarah Meiklejohn, and George
  Danezis.
\newblock Coconut: Threshold issuance selective disclosure credentials with
  applications to distributed ledgers.
\newblock {\em arXiv preprint arXiv:1802.07344}, 2018.

\bibitem{sasson2014zerocash}
Eli~Ben Sasson, Alessandro Chiesa, Christina Garman, Matthew Green, Ian Miers,
  Eran Tromer, and Madars Virza.
\newblock Zerocash: Decentralized anonymous payments from bitcoin.
\newblock In {\em 2014 IEEE symposium on security and privacy}, pages 459--474.
  IEEE, 2014.

\bibitem{groth2016size}
Jens Groth.
\newblock On the size of pairing-based non-interactive arguments.
\newblock In {\em Annual international conference on the theory and
  applications of cryptographic techniques}, pages 305--326. Springer, 2016.

\bibitem{ben2018scalable}
Eli Ben-Sasson, Iddo Bentov, Yinon Horesh, and Michael Riabzev.
\newblock Scalable, transparent, and post-quantum secure computational
  integrity.
\newblock {\em Cryptology ePrint Archive}, 2018.

\bibitem{bunz2018bulletproofs}
Benedikt B{\"u}nz, Jonathan Bootle, Dan Boneh, Andrew Poelstra, Pieter Wuille,
  and Greg Maxwell.
\newblock Bulletproofs: Short proofs for confidential transactions and more.
\newblock In {\em 2018 IEEE symposium on security and privacy (SP)}, pages
  315--334. IEEE, 2018.

\bibitem{zhang2020deco}
Fan Zhang, Deepak Maram, Harjasleen Malvai, Steven Goldfeder, and Ari Juels.
\newblock Deco: Liberating web data using decentralized oracles for tls.
\newblock In {\em Proceedings of the 2020 ACM SIGSAC Conference on Computer and
  Communications Security}, pages 1919--1938, 2020.

\bibitem{rosenberg2022texttt}
Michael Rosenberg, Jacob White, Christina Garman, and Ian Miers.
\newblock $ \texttt zk-creds$: Flexible anonymous credentials from zksnarks and
  existing identity infrastructure.
\newblock {\em Cryptology ePrint Archive}, 2022.

\bibitem{zhang2016town}
Fan Zhang, Ethan Cecchetti, Kyle Croman, Ari Juels, and Elaine Shi.
\newblock Town crier: An authenticated data feed for smart contracts.
\newblock In {\em Proceedings of the 2016 aCM sIGSAC conference on computer and
  communications security}, pages 270--282, 2016.

\bibitem{lee2013aggregating}
Kwangsu Lee, Dong~Hoon Lee, and Moti Yung.
\newblock Aggregating cl-signatures revisited: Extended functionality and
  better efficiency.
\newblock In {\em International Conference on Financial Cryptography and Data
  Security}, pages 171--188. Springer, 2013.

\bibitem{chase2014malleable}
Melissa Chase, Markulf Kohlweiss, Anna Lysyanskaya, and Sarah Meiklejohn.
\newblock Malleable signatures: New definitions and delegatable anonymous
  credentials.
\newblock In {\em 2014 IEEE 27th Computer Security Foundations Symposium},
  pages 199--213. IEEE, 2014.

\end{thebibliography}

\end{document}